\documentclass[aps,prl,twocolumn]{revtex4}
\usepackage[latin1]{inputenc}
\usepackage{amsmath}
\usepackage{amsfonts}
\usepackage{amssymb}
\usepackage[dvipdfmx]{graphicx}
\usepackage[dvipsnames]{xcolor}
\usepackage{bm}
\usepackage[a4paper,top=2cm,bottom=2cm,left=2cm,right=2cm]{geometry}

\newcommand{\dmiii}[1]{\textcolor{black}{#1}}
\newcommand{\dmii}[1]{\textcolor{black}{#1}}
\newcommand{\dmi}[1]{\textcolor{black}{#1}}

\begin{document}
	
\author{Davide Michieletto}
\thanks{davide.michieletto@ed.ac.uk}
\affiliation{School of Physics and Astronomy, University of Edinburgh, Peter Guthrie Tait Road, Edinburgh, EH9 3FD, United Kingdom}
\affiliation{MRC Human Genetics Unit, Institute of Genetics and Molecular Medicine, University of Edinburgh, North Crewe Rd, Edinburgh, EH4 2XU, United Kingdom}
%\affiliation{Centre for Mathematical Biology, and Department of Mathematical Sciences, University of Bath, North Rd, Bath, BA2 7AY, United Kingdom}

\author{Takahiro Sakaue}
\thanks{sakaue@phys.aoyama.ac.jp}
\affiliation{Department of Physics and Mathematics, Aoyama Gakuin University, 5-10-1 Fuchinobe, Chuo-ku, Sagamihara, Japan}
\affiliation{PRESTO, Japan Science and Technology Agency (JST), 4-1-8 Honcho Kawaguchi, Saitama 332-0012, Japan}

\title{Dynamical Entanglement and Cooperative Dynamics in Entangled Solutions of Ring and Linear Polymers}
	
\begin{abstract}
\textbf{Understanding how entanglements affect the behaviour of polymeric complex fluids is an open challenge in many fields. To elucidate the nature and consequence of entanglements in dense polymer solutions, we propose a novel method: a ``dynamical entanglement analysis'' (DEA) to extract spatio-temporal entanglement structures from the pair-wise displacement correlation of entangled chains. By applying this method to large-scale Molecular Dynamics simulations of linear and \dmi{unknotted, nonconcatenated} ring polymers, we find a strong and unexpected cooperative dynamics: the footprint of mutual \emph{entrainment} between entangled chains. We show that DEA is a powerful and sensitive probe that reveals previously unnoticed, and architecture-dependent, spatio-temporal structures of dynamical entanglement in polymeric solutions.  We also propose a mean-field approximation of our analysis which provides previously under-appreciated physical insights into the dynamics of generic entangled polymers. We envisage DEA will be useful to analyse the dynamical evolution of entanglements in generic polymeric systems such as blends and composites. 
} 
\end{abstract}

\maketitle

Entanglement is a fascinating and ubiquitous phenomenon in nature and yet a comprehensive microscopic theory of entanglement is still not established. While the tube and reptation models can approximate the material properties of entangled linear polymers~\cite{Gennes1979,Doi1988}, there are systems for which these theories do not apply. Among the most notable there are dense solution of ring polymers~\cite{Kapnistos2008,Halverson2011dynamics,Bras2011,Rosa2014}. Here, the global topological invariance of the system, i.e. the fact that \dmi{unknotted and nonconcatenated ring polymers must remain so} at all times, entails that the rings tend to collapse and to assume crumpled conformations which are not entangled with each other in a classical ``tube-like'' sense~\cite{Cates1986,Rubinstein1986,Rosa2014,Sakaue2011,Lang2012, Obukhov2014,Sakaue2016}. \dmi{Since rings do not have ends to diffuse, the reptation theory and its more modern extensions -- such as contour length fluctuations, constraint release or tube enlargement -- cannot be applied~\cite{McLeish2008,Doi_CLF_1983,Marrucci_tube_enlargement_1985}}. In light of this, a way to define and formalise entanglement that is also valid for topologically non-trivial polymers, such as rings, blends~\cite{Parisi2020} and higher order topological (or chimeric) polymers~\cite{Rosa2020tadpoles,Doi2015a,Uehara2016,Deguchi2017,Landuzzi2020a}, is highly needed. 

\begin{figure}[t]
\vspace{-0.6 cm}
	\centering
	\includegraphics[width=0.45\textwidth]{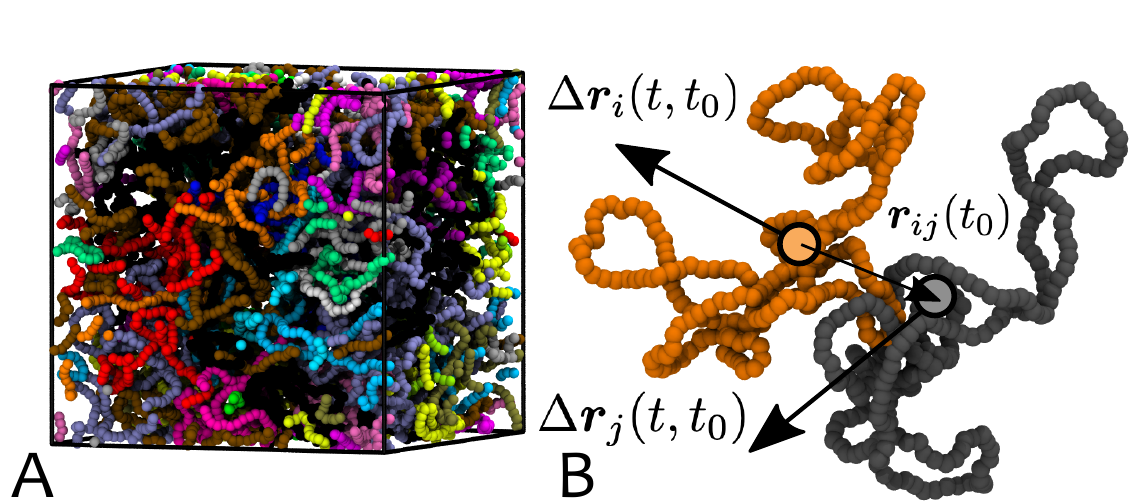}
	\caption{\textbf{A} Snapshot from MD simulations of a dense solution of $M=50$ ring polymers $N=256$ beads long within a box of size $L=50 \sigma$ with periodic boundaries. \textbf{B} The dynamical entanglement analysis (DEA) proposed in this work probes the correlation between displacements $\Delta\bm{r}_i(t,t_0)=\bm{r}_i(t_0+t)-\bm{r}_i(t_0)$ and $\Delta\bm{r}_j(t,t_0)=\bm{r}_j(t_0+t)-\bm{r}_j(t_0)$ of two polymers $i$ and $j$ at lag-time $t$ while recording their initial relative position $\bm{r}_{ij}(t_0)=\bm{r}_{j}(t_0) -\bm{r}_{i}(t_0)$. The correlation tensor $H(t,\bm{r})$ (Eq.~\eqref{eq:4point}) encodes the propensity of two polymers at relative position $\bm{r}$ to display correlated motion at timescale $t$. }
	\label{fig:4point}
			\vspace{-0.6 cm}
\end{figure}

\dmi{Here we propose to study the dynamical effect of topological constraints (TCs) by measuring the cooperative dynamics of entangled polymers in a complementary way with respect to other existing approaches, e.g. primitive path analysis~\cite{Everaers2004,Likhtman2014a,Likhtman2014b,Read2008} or constitutive equations~\cite{Read2008,Boudara2020}. Importantly, while most of the existing methods focus on either static entanglement structures or self-correlations (such as self mean squared displacement or stress relaxation) here we utilize the information of the correlated motion of entangled polymers in order to obtain information on the space- and time-dependent entanglement structures. We dub this method {\it dynamical entanglement analysis} (DEA)}. [\dmi{We also note that our approach to study cross-correlations (Fig.~\ref{fig:4point}) is different from that of earlier works~\cite{spiess1987,Cao2010}]}.
% and that the notion of cross-correlations is usually set aside in current standard analytical treatments (especially for ring polymers), where entanglements are effectively taken into account by introducing fixed obstacles.} 
%This reduction from a many-chain problem to a one-chain problem in a sea of obstacles -- or mean-field -- is at the heart of the tube and reptation models~\cite{Doi1988,Gennes1979}. 
%By removing the need for a ``sea of fixed obstacles'', the deviation of a test chain from a simple free motion must be manifested in the correlation of its dynamics with those of the neighbouring chains. This simple observation led us to the concept of DEA, in turn enabling the quantification of entanglement structures which evolve in space and time and that are encoded in the cooperative dynamics of polymers at corresponding time and spatial scales. 
%Importantly, while most of classical polymer physic models rely on single chain correlations, such as the mean square displacement (MSD) of a chain, our DEA can extract two-body correlations and thus shed light into spatio-temporal correlation patterns due to entanglement.

We apply DEA to large-scale Molecular Dynamics (MD) simulations of \dmi{entangled} systems of ring and linear polymers revealing qualitatively different spatio-temporal entanglement structures. Additionally, we link these numerical observations with phenomenological and mean-field theories which yield a renewed interpretation for the anomalous dynamics of entangled chains. 
%For rings, it provides a clear signature of an unique self-similar entanglement dynamics dominated by the non-concatenation topological invariance. Our analysis indicates that the number of entangled polymers decreases in time as a power-law for rings, a distinctive feature not seen in the case of linear chains. We argue that this feature may be at the heart of the different dynamics between these two topologies.

%{\it Results -- }	
\dmi{In practice, we consider $M$ polymers $N$ beads long in a cubic box of size $L^3$ with $\rho = M/L^3 $ denoting the concentration of chains and at monomer density $c=MN/L^3=0.1 \sigma^{-3}$ ($\sigma$ is the size of one bead). The polymers are semiflexible, i.e. have persistence length $l_p=5\sigma$, chosen to lower the typical entanglement length (found to be $N_e=40$ beads~\cite{Rosa2008}) and to promote inter-ring threading~\cite{Michieletto2016pnas}. A summarising table with the key parameters is given in the SI}. 
%\begin{table}[h!]
%\begin{tabular}{|c|c|c|c|c|c|c|}
% \hline 
% & \multicolumn{3}{|c|}{Ring} &  \multicolumn{3}{|c|}{Linear} \\ 
% \hline 
%N & $R_g^2$ $[\sigma^2]$ & L & M & $R_g^2$ $[\sigma^2]$ & L &  M \\
%\hline
%256 & 134.0 $\pm$ 0.2 & 50 & 50 & 349.1 $\pm$ 0.1 & 80 & 200 \\
%512 & 245.5 $\pm$ 0.4 & 60 & 40 & 723.6 $\pm$ 0.2 & 100.5 & 200 \\
%1024 & 397.4 $\pm$ 0.4 & 80 & 50 & 1354.6 $\pm$ 0.8 & 127 & 200 \\ 
% \hline 
% \end{tabular}
%\caption{\dmi{Length of the chains $N$, square radius of gyration $R_g^2$, size of the box $L$ and number of chains $M$ for systems of rings and linear chains. The entanglement length is $N_e=40$ beads~\cite{Everaers2004,Uchida2008b,Rosa2008,Rosa2013}. The choice of $L$ and $M$ are such that the size of box is about 4 times $R_g$ while keeping $c = \rho N$. }}
%\label{table}
%\end{table}

\begin{figure}[t!]
	\centering
	\hspace{-0.4 cm}
	\includegraphics[width=0.5\textwidth]{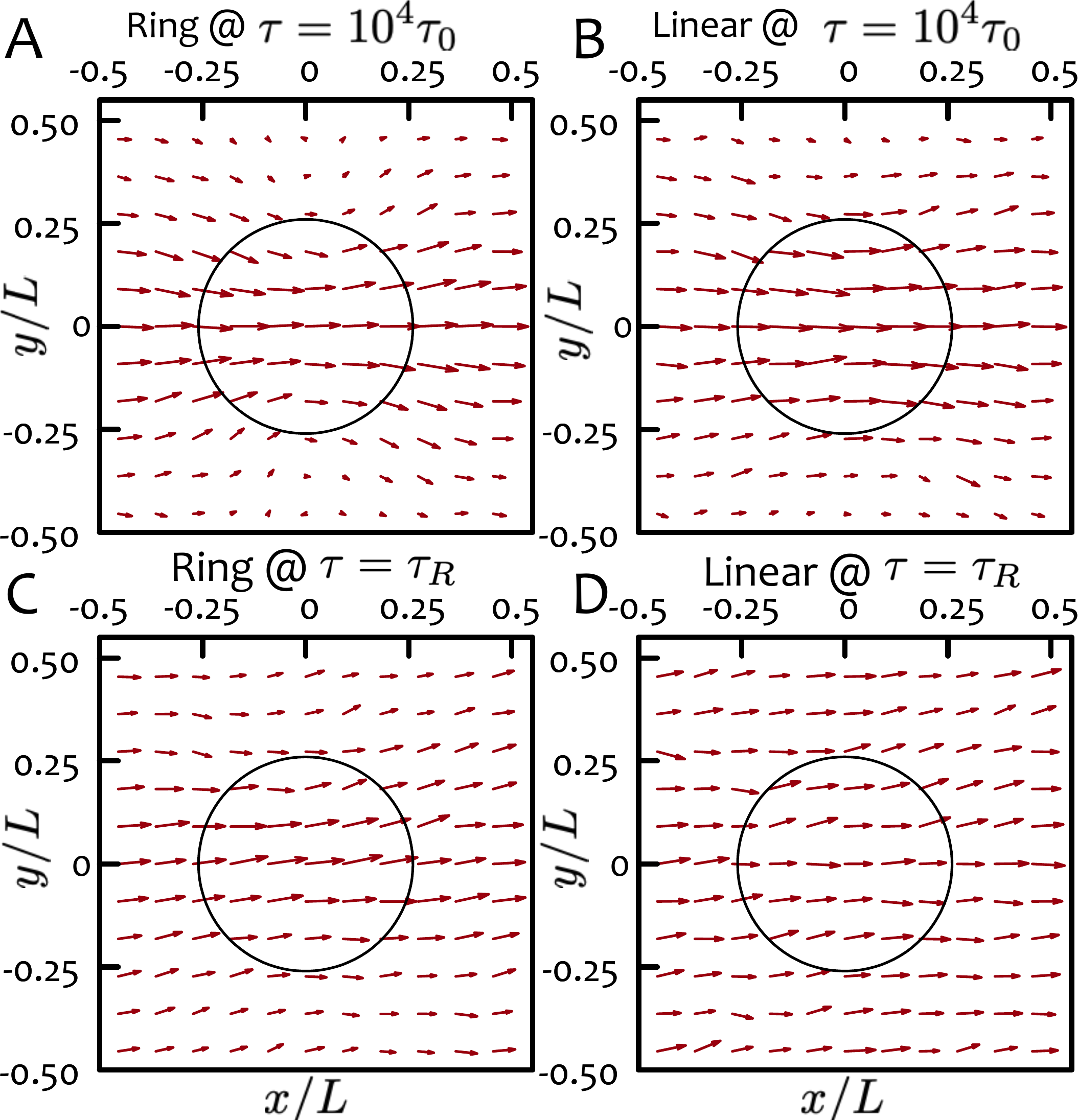}
	\caption{2D slice of the vector field $\bm{v}(t,\bm{r})= H(t,\bm{r}) \cdot \bm{e}_x$ for ring and linear polymers $N=512$ beads long at the lag-time (\textbf{A}-\textbf{B}) $t = 10^4 \tau_0$ and (\textbf{C}-\textbf{D}) $t = \tau_R=\tau_0 N^2 \simeq 26 \, 10^4 \tau_0$. Note that the vectors are scaled up or down for visualisation purposes: (\textbf{A}) $\times 2.5$ (\textbf{B}) $\times 20$ (\textbf{C}) $\times 0.125$ (\textbf{D}) $\times 1$. The circles are drawn using the average size $R_g$ of the polymers as radii \dmi{and note that the plots are normalised by the size of the box, which is different for ring and linear systems (see Table~I in SI)}. 
	}
	\label{fig:vfield_ringlin}
		\vspace{-0.4 cm}
\end{figure}

The correlation of displacement of different chains in the system can be computed as follows. Let $\bm{r}_i (t)$ be the position of centre of mass (CoM) of the $i$-th polymer at time $t$ and $\Delta \bm{r}_i(t, t_0) = \bm{r}_i(t_0+t)-\bm{r}_i(t_0)$ be its displacement during the lag time $t$. We define the displacement correlation tensor~\cite{Ooshida2016,Crocker2000} as 
\begin{equation}
\mathrm{H}_{\alpha \beta} (t,\bm{r}) \equiv \langle \Delta r_{i,\alpha}(t,t_0) \Delta r_{j,\beta}(t,t_0) \rangle_{\bm{r}} , \label{eq:4point}
\end{equation}
where the Greek indexes represent Cartesian components and the average $\langle \cdots \rangle_{\bm{r}}$ is intended over times $t_0$ and pairs of polymers $(i,j)$, which satisfy $\bm{r}_{ij}(t_0) \equiv \bm{r}_j(t_0) - \bm{r}_i(t_0)=\bm{r}$ (see Fig.~\ref{fig:4point}). We note that this formalism has been successfully employed in different contexts, e.g., in 2-point microrheology~\cite{Crocker2000} and to investigate dynamics near the glass transition in colloidal systems~\cite{Ooshida2016} but has never been directly applied to systems of polymers. As we show below, Eq.~\eqref{eq:4point} allows us to compute the correlation of the displacements at lagtime $t$ between polymers at relative position $\bm{r}$ (Fig.~\ref{fig:4point}) and provides more spatiotemporal information on entanglements with respect to conventional self-correlations.

%  the average is done over time and ensemble. 
%For isotropic system, $g(r)$, where $r=|\bm{r}|$, is known as the radial distribution function, and gives the probability of finding a ring at distance $r$ from any other ring. 
We compute $\mathrm{H}(t,\bm{r})$ on our systems of polymers and \dmi{choose not to subtract the motion of the CoM of the whole system (see below and SI for details)}. To visualize the correlation tensor, we impose a fictitious, arbitrary displacement $\bm{e}_x ={}^t(1,0,0)$, and plot a 2D slice of the resulting 3D vector field $\bm{v}(t,\bm{r})= \mathrm{H}(t,\bm{r}) \cdot \bm{e}_x$ for fixed lag-times. In Fig.~\ref{fig:vfield_ringlin} we show two examples at $t = 10^4 \tau_0$ and $\tau_R \equiv \tau_0 N^2$, i.e. the Rouse time (we identify the microscopic time scale $\tau_0$ with the Lennard-Jones (LJ) time $\tau_0=\tau_{LJ}=\sigma\sqrt{m/\epsilon}$, where $m$ is the mass and $\sigma$ the size of a bead and $\epsilon=k_BT$ the energy scale of the LJ potential).

Figure~\ref{fig:vfield_ringlin} captures the most important conceptual finding of this paper, i.e. that we observe a highly coordinated pattern of the vector fields representing correlation of displacements. \dmiii{Importantly, such a persistent, coordinated pattern is not observed in unentangled and phantom chains (see Figs.~S3 and S4 in SI)}.
\dmi{One should interpret these fields as the average displacement of a polymer at location $\bm{r}$ from a probe polymer placed at the origin and that has displaced a unit length horizontally $\tau$ timesteps earlier. The strong alignment of the correlation vectors should not be confused with a flow of the polymers but is rather a signature of strong correlated dynamics and the consequence of entanglement among chains. We connect this correlation with the notion of ``entrainment'' -- which amounts to a combination of steric and topological constraints between neighbouring polymers -- that effectively results in the mutual ``dragging'' of neighbours and correlated motion that can persist at least up to the Rouse time, $\tau_R \sim N^2$. We note that the vectors in Fig.~\ref{fig:vfield_ringlin} are scaled up/down for visualisation purposes (see caption) and that the rings display much stronger cooperation with respect linear chains as their correlation vectors are typically longer (before rescaling). } \dmiii{We stress that this is indication of rings being more entrained that linear chains of same length and that this is due to entanglement, as we observe no persistent coordinated correlation patterns in systems of phantom and unentangled polymers (see Figs.~S3, S4 in SI). Furthermore, removing the overall CoM motion in this case effectively corresponds to constraining the overall flux of the correlation vector field over the simulation box to be almost zero (actually about $1/M$ as we show in the SI).}  

In order to further quantify this cooperative dynamics, we define the scalar quantity
\begin{eqnarray}
\chi(t, r) =\dfrac{{\rm Tr}[ H(t, r)]}{g_3(t)}, \label{eq:chi4}
\end{eqnarray}
which measures the degree of correlation during the time scale $t$ between a pair of polymerss that are initially separated by a distance $r=|\bm{r}|$. In Eq.~\eqref{eq:chi4}, $g_3(t)=\langle( \bm{r}_{CM}(t_0+t) - \bm{r}_{CM}(t_0))^2 \rangle$ is the mean squared displacement (MSD) of the polymers' CoM (or mean-squared self-displacement). [Note that ${\rm Tr}[ H(t, \bm{r})]$ is expected to depend only the separation $r = |\bm{r}|$ for homogeneous and isotropic system as the one considered here.] To elucidate the physical meaning of $\chi(t, r)$, imagine that we start to apply a force $f$ in $x$ direction at $t=0$ only to the polymer $i$ at the origin. This force will cause an average displacement of polymer $i$ that can be computed as $ \langle \Delta r_{i, x}(t) \rangle = ft/\Gamma(t) = f g_3(t)/6k_BT$, where $\Gamma(t) =k_BT/D(t)$ is the effective friction, and $6D(t)t=g_3(t)$. \dmi{At the same time, if chain $i$ is entangled with chain $j$, it may cause the motion of chain $j$ initially located at $\bm{r}_j(t_0)=\bm{r}$.} Its average displacement can be calculated in a similar way as done for the probe chain: 
\begin{eqnarray}
 \langle \Delta r_{j, \alpha}(t) \rangle_{\bm{r}} =
  \frac{H_{\alpha \beta}(t, \bm{r}) f_{\beta}} {2k_BT}  \, ,
\label{Delta_r_j}
\end{eqnarray}
where we generalise the mean-squared self-displacement in time ($g_3$) to include pair-wise correlation in space-time and along different directions via the time and space dependent tensor $H_{\alpha \beta}(t,\bm{r})$. For an isotropic system, we can take the angular average, leaving only the component along the direction of the force in Eq.~(\ref{Delta_r_j}), i.e.
$ \langle \Delta r_{j, x}(t) \rangle_{r} =  \frac{f} {6k_BT}  \, {\rm Tr}[ H(t, r)]$. In the limit of perfect cooperativity $H_{\alpha \beta}(t,r)=2 \delta_{\alpha \beta} D(t) t$ and the previous equation reduces to that for the probe chain. 

In light of this, $\chi(t, r)$ can be seen as the ratio between the average displacement induced on polymer $j$ and the average self-displacement of the probe $i$ at time $t$ and conditional to the fact that $|\bm{r}_i(t_0)-\bm{r}_j(t_0)|=r$. \dmi{As such, $\chi(r,t)$ is bound to take values between 0 and 1, and may be thought as the fraction of monomers of polymer $j$ that are effectively ``dragged'' -- or entrained -- by the motion of the (entangled) polymer $i$ during the time scale $t$.} Thus, the DEA naturally yields a quantity that has a physically appealing and intuitive meaning, that of how many monomers of polymer $j$ at position $r$ are entrained by $i$, $t$ timesteps after that it has moved.

At long length- and time-scales, we expect the hydrodynamic behaviour with macroscopic viscosity $\eta_b$ to dominate. In this regime $dH_{\alpha \beta}(t,\bm{r})/dt =(k_BT/8 \pi \eta_b r) (\delta_{\alpha \beta} + {\hat r}_{\alpha}{\hat r}_{\beta})$ is the Oseen tensor and $g_3(t) \simeq (k_BT/6 \pi \eta_b R_g) t$. In this limit, $\chi(t, r)$ decays as $\sim R_g/r$. On the other hand, at length scales $r \lesssim R_g$, a cooperative motion will be predominantly caused by entanglements between polymers and $\chi(t,r)$ is thus expected to be a sensitive measure of the dynamically evolving entanglement.

\begin{figure}[t!]
	\centering
	\hspace{-0.4cm}
	\includegraphics[width=0.5\textwidth]{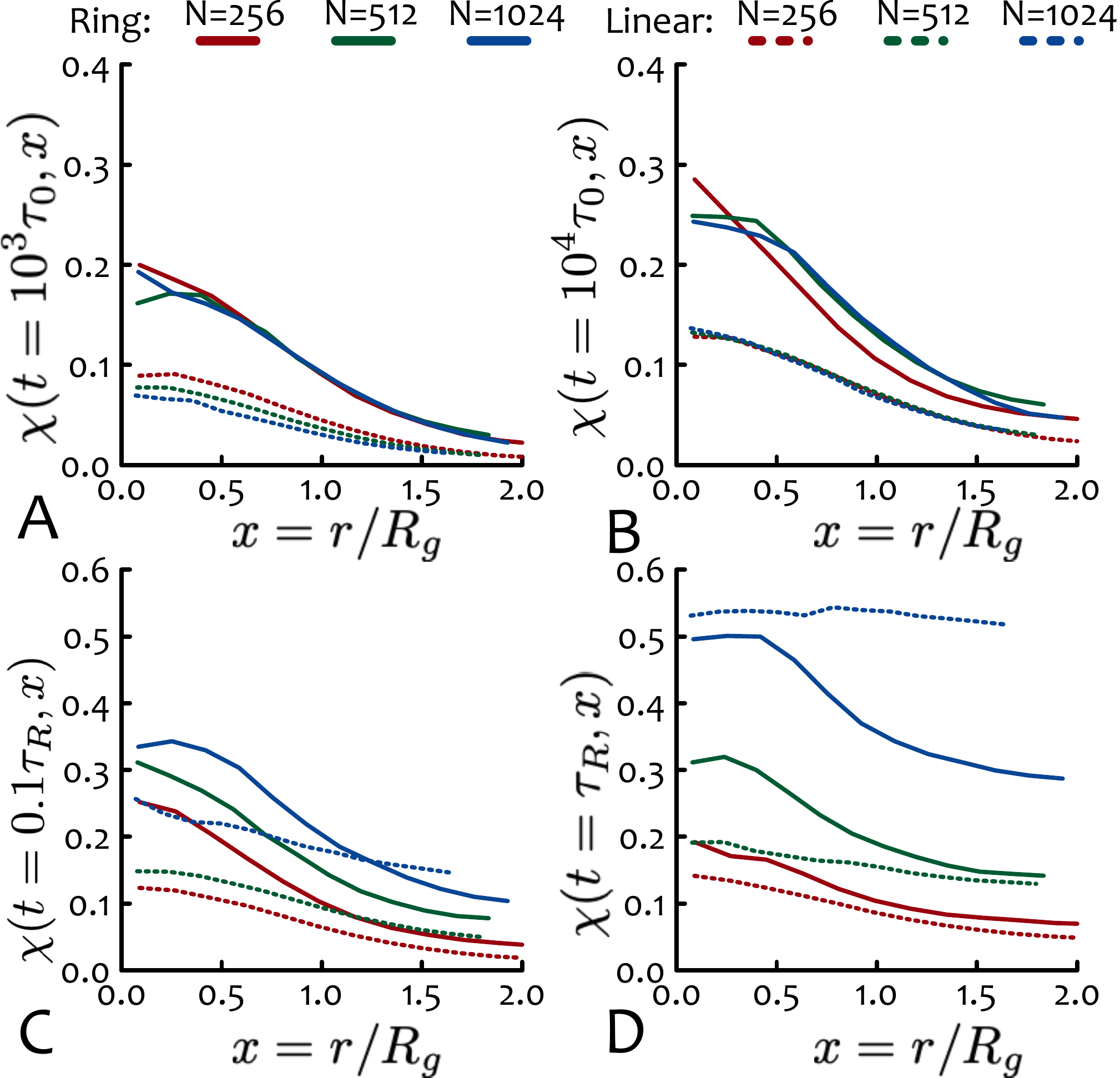}
	\caption{Spatio-temporal evolution of dynamical cooperativity $\chi(t ,r/R_g)$. This figure shows the spatial profile of $\chi(t,r)$ at fixed lag-times: (\textbf{A}) $t= 10^3 \, \tau_0$, (\textbf{B}) $t =10^4  \tau_0$, \textbf{(C)} $t= 0.1 \tau_R= 0.1 \tau_0 N^2$ and \textbf{(D)} $t= \tau_R = \tau_0 N^2$. Notice that the lag-times in \textbf{(C)} and \textbf{(D)} are $N$ dependent, hence the larger $\chi(t, r)$ values for longer chains.
	%Notice that at short times the correlation is larger for rings than for linear chains, but the decay is steeper for rings.  
	The maximum length-scale here is $L/2$ due to the periodic boundary conditions and is equivalent to $2R_g$ in our simulations. 
	%\dmi{Comment: in (B), can we also see the $\chi$ data in intermediate time range, say, $t = 10^4$ or $10^5 \tau_0$, or $t = 0.1 \tau_R$. Compared to the first two cases, the time in the last case $t = 0.1 \tau_R$ is different for chains with different length, so we need some caution for that. }}
	}
	%(\textbf{C})-(\textbf{D}) Show the temporal evolution of $\chi(t,r)$ at fixed distance (\textbf{C}) $r= 0$ and (\textbf{D}) $r=R_g$.
	\vspace{-0.4 cm}
	\label{fig:chi4}
\end{figure}

In Fig.~\ref{fig:chi4}, we show $\chi(t,r)$ at different lag-times $t$ and as function of $r$. One can notice that at short lag-times (Fig.~\ref{fig:chi4}A,B) $\chi (t,r)$ follows an architecture-dependent but length-independent master curve. \dmi{In line with the magnitude of the vector field in Fig.~\ref{fig:vfield_ringlin}, $\chi(t,r)$ also takes larger values for rings than for linear chains, with the latter catching up the former with time.} The steep spatial gradient indicates that rings with overlapping CoMs (hence likely interpenetrating, or threading~\cite{Michieletto2014acs,Michieletto2016pnas,Lee2015,Smrek2016}) are more correlated than distant ones. This spatial dependence is instead weaker for linear chains, indicating a more uniformly distributed entanglement structure over the whole contour of the chain, in qualitatively agreement with the picture of the tube model. In other words, the structure of entanglements between neighbouring chains is qualitatively different between ring and linear chains, and this difference is mirrored in weaker/stronger correlated motion shown in Fig.~\ref{eq:chi4}. \dmiii{We also stress that for phantom and unentangled chains, $\chi$ remains close to zero indicating that, as expected, there is no correlation at any spatial or temporal scales for non-entangled chains (see SI, Figs.~S3 and S4).}

%The cooperative motion at this time scale is likely due to the mutual entrainment of chains caused by their mutual entanglement. Indeed, Fig.~\ref{fig:chi4}A suggests that two rings with their centres of mass positioned nearby are strongly constrained by their relative motions, but such dynamical entanglement decreases rather sharply with their separation. On the contrary, linear chains display shallower gradients in $\chi(t,r)$ indicating a weaker spatial dependence of the entanglement which can thus be interpreted as more uniformly distributed over the whole contour of the chain.

%At larger times the gradient is shallow for both topologies and the correlations displayed by linear polymer are at least as strong as those for rings. This suggests that as rings sample their surrounding space, they re-distribute the dragging across its contour length and among the neighbouring chains ultimately achieving a weaker dependence on the distance from the centre of mass.   

%In order to obtain a more quantitative connection between the inter-chain displacement correlation and the mutual dragging due to \emph{dynamical entanglement}, we calculate the spatial average of $\chi(t,r)$

\begin{figure}[t!]
	\centering
	\hspace{-0.4cm}
	\includegraphics[width=0.5\textwidth]{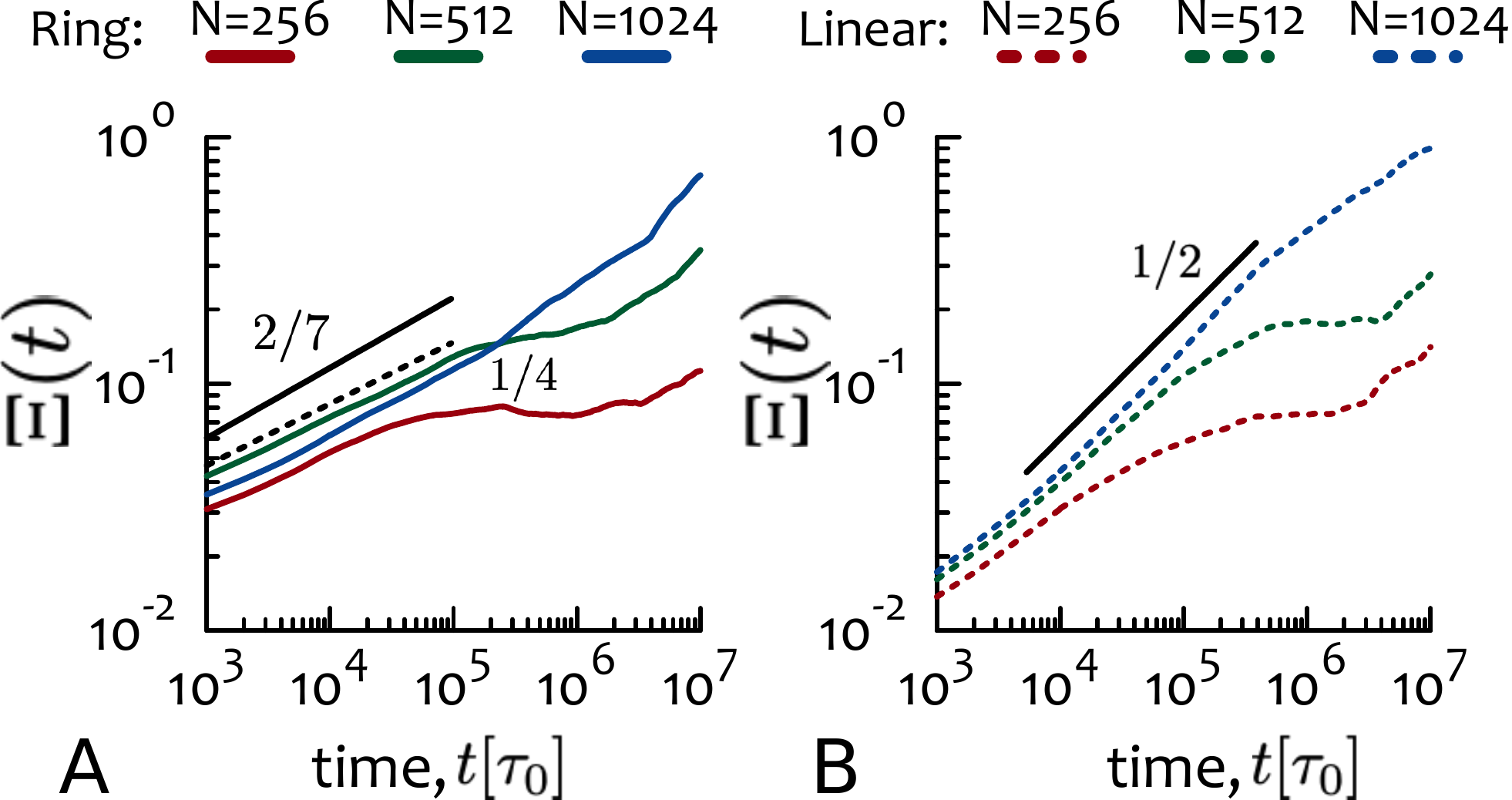}
	\caption{Spatial average of the dynamical correlation, $\Xi(t)$, for \textbf{(A)} rings and \textbf{(B)} linear chains. The scaling for rings, $\sim t^{2/7}$ and $\sim t^{1/4}$ are obtained from Eq.\eqref{eq:scaling} with $d_f=3$ (crumpled globule) and $d_f=5/2$ (Cates-Deutsch crossover~\cite{Cates1986,Halverson2011statics}, see SI), respectively. For linear chains our calculation agrees with the classical reptation picture which predicts $\Xi(t)\sim t^{1/2}$ (see Eq.\eqref{eq:scaling}).
	%We note that the systems with $N=256$ is poorly entangled.
	%Solid lines indicate rings whereas dashed lines correspond to linear chains. Black dot-dashed and solid lines mark scaling $1/4$ and $1/2$ respectively and serve as guide for the eye. 
	%In (\textbf{B}) the relaxation time $\tau_c$ is computed from the MSD as $g_3(\tau_c) \equiv R_g^2$ for both ring and linear chains (see SM for plots of $g_3(t)$). 
	%\dmi{Comment: Why $\Xi(t)$, which is the spatial average of $\chi(t, r)$, becomes larger than unity, although $\chi(t, r) < 1$? Another issue: I think only figure A is enough here, because there is no theoretical ground to expect the scaling form $\Xi \sim (t/tau_d)^{p}$.  For figure A, is it possible to replace the vertical axis from $\Xi$ to $\Xi P$, where $P$ is defined in the text? With this, I expect a better collapse; see the added equation for $\Xi$ in SM. } }
	}
		\vspace{-0.2 cm}
	\label{fig:S}
\end{figure}

To connect DEA with classic theories, we develop a mean-field approximation of our analysis in line with the mean-field approach of the tube theory. We calculate the spatial average of $\chi(t,r)$ as follows
\begin{equation}
\Xi(t) \equiv \frac{4 \pi \rho \int_0^{r^{*}}r^2 g(r) \chi(t,r) \ dr}{  4 \pi \rho \int_0^{r^{*}}r^2 g(r)  \ dr  }
\label{eq:intChi}
\end{equation}
where $g(r)$ is the radial distribution function of the polymers' centres of mass, and we take $r^{*}=L/2$ due to the periodic boundary condition. This quantity, plotted in Fig.~\ref{fig:S}, exhibits a power-law behavior in the short-medium time scale.
 %$g(\bm{r})= (\rho )^{-1} \sum_{i \neq j}\left\langle \delta((\bm{r}_{j}(t_0) -\bm{r}_{i}(t_0))- \bm{r}) \right\rangle$
The scaling of $\Xi(t)$ can be understood in light of the dynamical entanglement picture: since each chain has a number $P = 4 \pi \rho \int_0^{r^{*}}r^2 g(r) \ dr$ of surrounding chains, i.e. those in the spherical volume of radius $r^{*}$, the number of monomers dragged by the motion of $i$-th ring is $\sim P N \Xi(t)$ and this provides an estimate for the effective friction experienced by each chain as $\Gamma(t) \simeq \gamma_0 P N \Xi(t)$ with $\gamma_0$ being a segment friction. 

This effective friction on the global motion of the polymer can be compared with the theoretical predictions of the reptation theory for linear chains~\cite{Doi1988} and more recent theories for rings~\cite{Ge2016}. For linear polymers, the simplest reptation theory suggests~\cite{Doi1988} that $g_3(t) \sim \sigma^2 (N_e^2/N)(t/\tau_e)^{1/2}$ for $\tau_e < t < \tau_{R}$, where $N_e \simeq 40 $~\cite{Rosa2014,Michieletto2016pnas} is the entanglement length and $\tau_e \sim \tau_0 N_e^2$ is the relaxation time for the entanglement lengthscale. For non-concatenated ring polymers, the conformation and the dynamics are self-similar~\cite{Rubinstein1986,Ge2016,Halverson2011dynamics,Rosa2014} for chain sections that are longer than $N_e$ and on time scales $\tau_e < t < \tau_c$, where $\tau_c$ is the longest conformational relaxation time. According to the loopy-globule model~\cite{Ge2016}, $\tau_c \sim \tau_e (N/N_e)^{(2d_f+1)/d_f}$, which may be thought as an analogue of the Rouse time in the linear chain counterpart ($d_f$ denotes the fractal dimension of the ring conformation at the scale larger than the entanglement length). As shown in SI, the MSD of mass center turns out to be $g_3(t) \sim \sigma^2 (N_e^2/N) (t/\tau_e)^{(d_f+2)/(2d_f+1)}$ for $\tau_e < t < \tau_c$. With the help of the Einstein relation, these considerations lead to the effective friction $\Gamma(t) \simeq (k_BT/g_3(t))t$, which scales as
 \begin{eqnarray}\label{eq:scaling}
\Gamma(t) \sim \Xi(t) \sim
\left\{
\begin{array}{ll}
t^{1/2}  & (\tau_e < t < \tau_R) \quad {\rm linear} \\
t^{2/7}  & ( \tau_e < t < \tau_c) \quad {\rm ring}\  d_f=3
\label{Gamma}
\end{array}
\right.
\end{eqnarray}
in good agreement with the numerical observation in Fig.~\ref{fig:S}. Note the exponent for ring becomes $1/4$ if we adopt the effective fractal dimension $d_f=2.5$ which is more appropriate for rings with intermediate length~\cite{Cates1986,Halverson2011statics} (see also SI). 
%(see SM for more complete forms of $\Gamma(t)$ and $\Xi(t)$ from the above discussion).

\begin{figure}[t!]
	\centering
	\hspace{-0.4 cm}
	\includegraphics[width=0.5\textwidth]{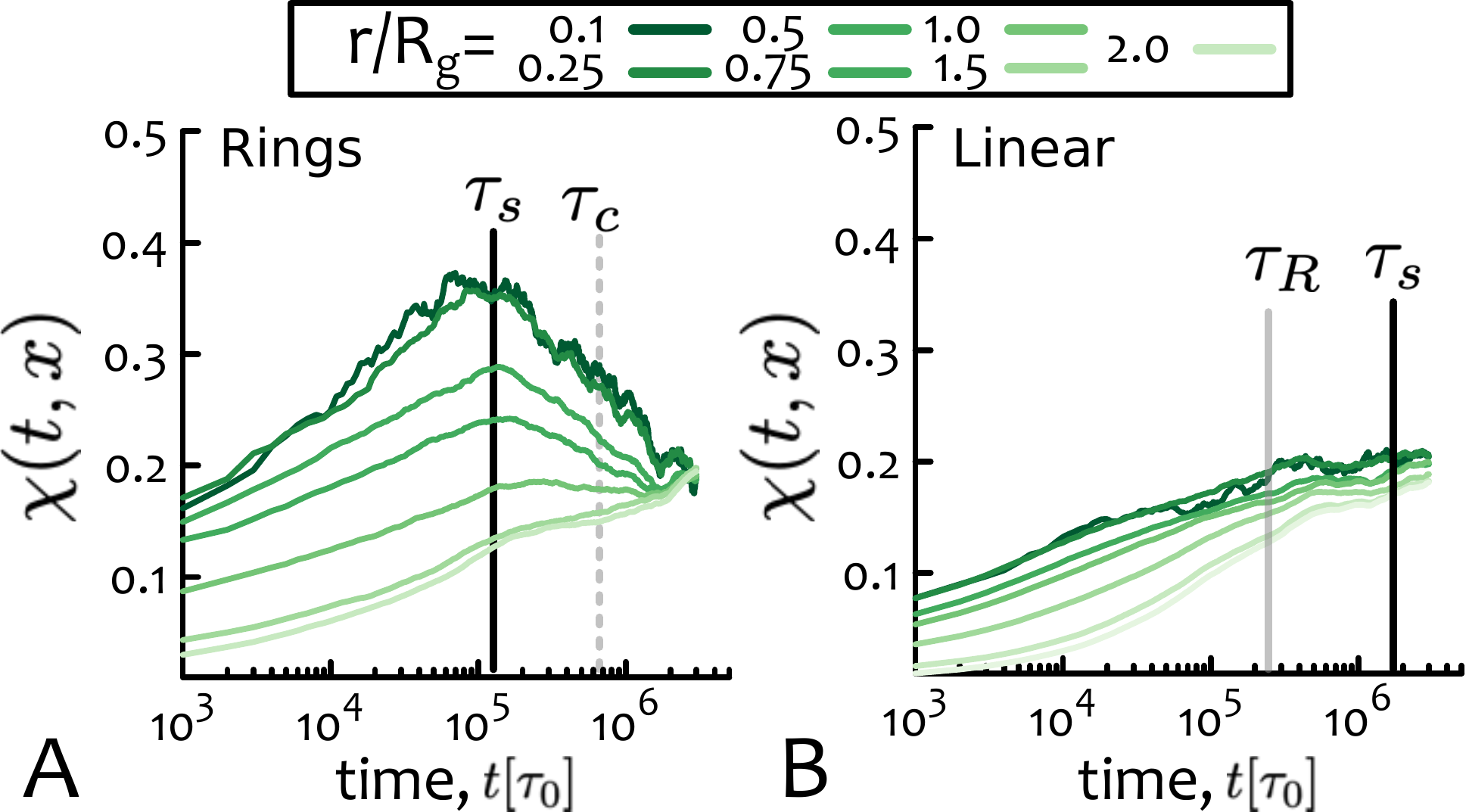}
	\caption{Displacement correlations display a topology-dependent behaviour. We compare $\chi(t,r)$ (Eq.~\eqref{eq:chi4}) for (\textbf{A}) ring and (\textbf{B}) linear chains against time for increasing distances $r/R_g$ and for fixed polymer length $N=512$. In the figure, $\tau_s$ represents the self-diffusion time of the polymers, i.e. $g_3(\tau_s)\equiv R_g^2$ (see SI for $g_3(t)$ curves), $\tau_c = \tau_0 N_e^{-1/3}N^{7/3}$ the longest conformational relaxation time for rings and $\tau_R = \tau_0 N^2$ the Rouse time.  Notice the non-monotonic behaviour of $\chi(t,r)$ at short length-scales that is peculiar for rings and indicating that the number of entanglements changes in time in systems of rings.
%	\dmi{Can you modify the grey line for time scale here as follows? Draw the line at $t=t_c= \tau_0 N_e^{-1/3}N^{7/3}$ (see equation in text) for ring and $t=\tau_R = \tau_0 N^2$ for linear. In addition, another line at $t=\tau_d$ for linear, which is the current line in Fig. (B) I think. In addition, I'm a bit skeptical about the diffusion time $\tau_d$ for ring, which seem too short. In fact, if we look Fig. 6 A in SM, the ring CM keeps subdiffusion behavior in much longer time scale.}  }
}
	\label{fig:chi4_vs_time}
			\vspace{-0.4 cm}
\end{figure}

A further insight on the dynamical entanglement might be obtained by the following consideration for linear chains: there are $X_0 \equiv N/N_e$ polymers, on average, entangled with the $i$-th chain. In our picture, each entangled chain is entrained by the motion of the $i$-th chain, but such a response does not take place immediately but progressively and following the propagation of the tension from the entanglement points~\cite{Rubinsteinbook,Sakaue2016tension}. For linear chains, the Rouse model predicts that $n(t) \sim N_e (t/\tau_e)^{1/2}$ monomers (counting from each entanglement point) move coherently with the $i$-th chain and thus contribute to the friction at time scale $\tau_e < t < \tau_R$. This leads to the effective friction
\begin{eqnarray}
\Gamma_{\rm linear}(t) \sim \gamma_0 N_e \left(\frac{n(t)}{N_e}\right) \times X_0 \qquad  (\tau_e < t < \tau_R)  \, . 
\end{eqnarray}
Besides being in agreement with Eq.~(\ref{Gamma}), the above discussion provides a physically appealing picture for the well-established $g_3(t) \sim t^{1/2}$ scaling for the polymers' CoM motion in systems of linear chains. Now, anticipating a similar argument for ring polymers, we rewrite this effective friction as
\begin{eqnarray}
\Gamma_{\rm ring}(t) \sim \gamma_0 N_e \left(\frac{n(t)}{N_e}\right) \times X(t) \qquad  ( \tau_e < t < \tau_c) 
\end{eqnarray}
and require it to match Eq.~(\ref{Gamma}), where $n(t)$ and $\Gamma_{\rm ring}(t)$ are obtained from the loopy-globule model (see SI). Remarkably, we find that the number of entangled chains is now \dmi{time-scale dependent}  and it takes the form
\begin{equation}
X(t) = X_0 \left( \frac{n(t)}{N_e}\right)^{-1/d_f} \sim X_0 \left( \frac{t}{\tau_e}\right)^{-1/(2d_f+1)} \, .  \label{eq:dynent}
\end{equation}
The number of entangled rings will thus be reduced from $X_0 = N/N_e$ at $t = \tau_e$ to $(N/N_e)^{1-(1/d_f)}$ at $t=\tau_c$. Such a feature is absent in the entangled linear polymers for which they have, on average, the same number of entangled chains at all times. Thus the predicted power-law scaling for the number of entangled chains represents a unique feature in ring polymers system,  which is akin to a tube dilation effect proposed in the loopy globule model~\cite{Ge2016}. Note that in solutions in which $N/N_e$ exceeds the coordination number (or the overlap parameter) $\sim R^3/N\sigma^3$, which scales as $\sim N^{1/2}$ for linear chains but takes constant value of about $15-20$ for rings~\cite{Kavassalis1987a,Rosa2014}, the above discussion may require some modification~\cite{Ajdari1995}.

{\it Discussion -- } 
Using DEA, we find that a chain mobility is dictated by entanglement-driven \emph{entrainment}, and that the number of such entrained segments increases in time due to the propagation of entanglements up to the Rouse time $\tau_R$ for linear chains and the conformational relaxation time $\tau_c$ for rings. Beyond these time scales, our current data does not allow a quantitative discussion because of the insufficient statistics, but it is certainly interesting to explore the behavior of displacement correlation in such a longer time scale or larger systems. As a preliminary discussion, we show in Fig.~\ref{fig:chi4_vs_time}  the spatio-temporal evolution of $\chi(t, r)$, now plotting several curves at fixed length scale $r$ and as a function of time. 
%Several marked differenced are evident between ring and linear systems. First, as mentioned previously, the spatial decay of the correlation is much steeper for rings. 
%Second, while for linear chains $\chi(t, r)$ keep increasing at all length scales even after $\tau_R$, they display a decrease at $\tau_c$ for short length scales.
Several marked differences between ring and linear chains are evident but the most remarkable is the following: for linear chains, $\chi(t,r)$ increases until the self-diffusion time $\tau_s$ (defined as $g_3(\tau_s) = R_g^2$) suggesting that this coincides with the disentanglement time~\cite{Gennes1979,Doi1988,Rubinsteinbook} at which the correlated motion by mutual entanglement/entrainment disappears. On the contrary, for ring polymers we see that $\chi(t,r)$ is non-monotonic and its decay at short length scales indicates -- in agreement with our model discussed above -- that the number of entangled rings $X(t)$ decreases in \dmi{time} (Eq.~(\ref{eq:dynent})). At the same time, the dynamical cooperativity at large length scales displays a monotonic increase even at times larger than the self-diffusion time, i.e. $t > \tau_s$, which can be understood as due to the remaining $X(\tau_s)$ entangled chain at $\tau_s$ yet to be released from the entrainment. Further, we observe remnants of increasing dynamical entanglement at $\tau_c$ and argue that these are fully consistent with, and indeed rationalise, previous results showing that the time to enter free diffusion is considerably longer than the self-diffusion and conformational relaxation times ($\sim \tau_c$) in the dense solution of rings~\cite{Halverson2011dynamics,Michieletto2016pnas,Michieletto2017prl,Lee2015}.
%\dmii{A recent study has shown that such a slow diffusion phenomenology is well described by invoking mandatory cooperative structural rearrangement~\cite{Sakaue2018}. This may be related to the present argument that there are still $(N/N_e)^{1-(1/d_f)}$ rings entangled at $t=\tau_c$. Does this disentanglement process, if present, involve the hypothesized cooperative structural rearrangement and what is its relationship with the (un)threading dynamics between rings? Closing in on this speculation requires further extensive simulations and theoretical investigations.}
A possible scenario for such a slow diffusion phenomenology is to invoke a mandatory cooperative structural rearrangement of rings~\cite{Sakaue2018}, which may be related to the residual entanglement $X(\tau_c)$. \dmi{We also argue that in the case of artificially frozen polymers~\cite{Michieletto2016pnas,Michieletto2017prl} or partially active ones~\cite{Smrek2020}, the entanglements that induce the entrainment found via DEA here may be the ones that induce the topological glass state.}

%At the same time, the systems with polymers of length $N=256$ are poorly entangled and therefore should be seen as a ``negative control'', i.e. representing systems with weak or absent dynamical entanglement.
%According to our current knowledge on linear polymers, the chain goes into a disentanglement stage by reptating along the tube. It is known that a characteristic time scale for the disentanglement roughly corresponds to the diffusion time $\tau_d$. For ring polymers, the situation is less clear. 

%To summarize, the results in Fig.~\ref{fig:chi4_vs_time} suggest a scenario for linear polymers in which the decay of dynamical correlation proceeds with the disentanglement. For rings, the situation is more puzzling, but we argue it is connected with the numerical observations that the diffusion time is noticeably longer than conformational relaxation. 

{\it Conclusions -- } 
We have shown that the DEA (Eq.~\eqref{eq:4point}) can provide rich information on the spatial and temporal evolution of architecture-specific entanglements in polymer solutions. The proposed approach can be applied to various systems, e.g., from solutions of active polymers to chromosomes in vivo. For instance, a direct experimental realization of DEA may be feasible using fluorescently labelled actin or DNA~\cite{Abadi2018}; while at present most of the experimental observables focus on self-correlations, it may be interesting to look at cross-correlation of differently-tagged molecules. 

\dmi{Furthermore, and additionally to the motion of the CoMs, it would possible to analyse the interchain dynamic correlation at the segment scale, or perhaps more interestingly over an entanglement length scale; this may provide additional insights for instance clarifying finite size effects on DEA, which has not been done here.} \dmii{We also note that the surprisingly long and spatially extended cooperativity observed using DEA and the notion of \emph{entrainment}, may explain the discrepancies of experiments with theories for linear~\cite{Ylitalo1991} and ring~\cite{Kapnistos2008} polymers and their long subdiffusive regime compared with their self-diffusion time.}
%Additionally to the motion of the CoMs, it is possible to analyse the correlation of segment dynamics which may provide additional insights. 

 \dmii{Finally, we highlight that our DEA can be applied to any polymeric system and we envisage interesting outcomes on entangled blends~\cite{Parisi2020}, composites~\cite{Michieletto2019softmatter,Fitzpatrick2018} or chimeric~\cite{Rosa2020tadpoles} polymer systems}.  

%By comparing the mean-field picture of our analysis and the phenomenological models, we have provided a renewed interpretation for the anomalous dynamics of polymer center of mass. This naturally leads to the finding that the number of entanglement chains decreases with time for the ring polymers.
%Such a view is supported by the comparison of $\Gamma(t)$ and $\Xi(t)$: the former being the friction coefficient of the centre of mass evaluated by theoretical models and the latter the spatial average of $\chi(t,r)$, numerically calculated here (Fig.~\ref{fig:S}). Additionally, by inspecting $\chi(t,r)$ we report markedly distinct spatio-temporal behaviours of dynamical entanglements and qualitatively different modes of correlation relaxation for linear and ring polymers (Fig.~\ref{fig:chi4_vs_time}).

\paragraph{Acknowledgements.} This work was in part supported by JSPS KAKENHI (No. JP18H05529) from MEXT, Japan, and JST, PRESTO (JPMJPR16N5). DM is supported by the Leverhulme Trust through an Early Career Fellowship (ECF-2019-088). TS thanks T. Ooshida for fruitful discussion on displacement correlation analysis. The authors would like to acknowledge networking support by the COST Action CA17139.\\

\bibliography{Polymers,library}

\end{document}